\documentclass[aps,prl,twocolumn,groupedaddress,superscriptaddress]{revtex4-2}
\usepackage{graphicx}
\usepackage{amsmath}
\usepackage{epstopdf}
\usepackage{float}
\usepackage{color}
\usepackage{appendix}

\usepackage{multirow}
\usepackage{booktabs}
\usepackage{longtable}

\begin{document}
	\title{Observation of a broad state-to-state spin-exchange collision\\ 
		 near a $p$-wave Feshbach resonances of $^6$Li atoms}
	
		\author{Shuai Peng}
		\thanks{These authors contributed equally to this work.}
		\affiliation{School of Physics and Astronomy, Sun Yat-Sen University, Zhuhai 519082, China}
		
		\author{Tangqian Shu}
		\thanks{These authors contributed equally to this work.}
		\affiliation{School of Physics and Astronomy, Sun Yat-Sen University, Zhuhai 519082, China}
		
		\author{Bowen Si}
		\thanks{These authors contributed equally to this work.}
		\affiliation{School of Physics, Dalian University of Technology, Dalian 116024, China}
		
		\author{Sijia Peng}
		\affiliation{School of Physics and Astronomy, Sun Yat-Sen University, Zhuhai 519082, China}
		
		\author{Yixin Guo}
		\affiliation{School of Physics and Astronomy, Sun Yat-Sen University, Zhuhai 519082, China}
		
		\author{Yongchang Han}
		\affiliation{School of Physics, Dalian University of Technology, Dalian 116024, China}
		
		\author{Jiaming Li}
		\email[]{lijiam29@mail.sysu.edu.cn}
		\affiliation{School of Physics and Astronomy, Sun Yat-Sen University, Zhuhai 519082, China}
		\affiliation{Center of Quantum Information Technology, Shenzhen Research Institute of Sun Yat-Sen University, Shenzhen 518087 China}
		\affiliation{Quantum Science Center of Guangdong-HongKong-Macao Greater Bay Area, Shenzhen 518048, China}
		
		\author{Gaoren Wang}
		\email[]{gaoren.wang@dlut.edu.cn}
		\affiliation{School of Physics, Dalian University of Technology, Dalian 116024, China}
		
		\author{Le Luo}
		\email[]{luole5@mail.sysu.edu.cn}
		\affiliation{School of Physics and Astronomy, Sun Yat-Sen University, Zhuhai 519082, China}
		\affiliation{Center of Quantum Information Technology, Shenzhen Research Institute of Sun Yat-Sen University, Shenzhen 518087 China}
		\affiliation{Quantum Science Center of Guangdong-HongKong-Macao Greater Bay Area, Shenzhen 518048, China}
		
	
	
	\date{\today}

	\begin{abstract}
	The study of state-to-state spin-exchange collisions in the vicinity of $p$-wave Feshbach resonances offer great opportunities to explore many-body interactions and novel quantum phases.  Here, we report the observation of a spin-exchange collision near a $p$-wave Feshbach resonance within a mixture of the lowest and third-lowest hyperfine states of $^6$Li atoms. The spin-exchange interaction is observed over a range of ten gausses and produces a pair of atoms in the second-lowest hyperfine states that are captured by a deep optical dipole trap. We apply a coupled-channel method to calculate the scattering properties of this system. We find that the $p$-wave resonance exhibits a low inelastic collision rate and a broad resonance profile, which is due to the modification by the accompanying spin-exchange collisions.  These findings open up new possibilities for the creation of long-lived, strongly interacting  $p$-wave Fermi gases.

	\end{abstract}
	\maketitle
	
	
Feshbach resonance is a phenomenon in atomic and molecular physics where the scattering properties of two colliding particles are significantly altered due to the coupling between different quantum states~\cite{Chin2010RMP82.1225}. This coupling arises when the energy of the colliding particles is tuned to match the energy of a bound state of the system, leading to a resonance effect. Feshbach resonances play a crucial role by enhancing the scattering properties of the atoms involved~\cite{OHara2002Science298.5601.2179, Li2018PRL120.193402,Park2023Nature614.54, Su2022PRL129.033401}. Specifically, Feshbach resonances can increase the elastic scattering cross-section, which is beneficial for processes like evaporative cooling, as it helps in achieving thermal equilibrium more efficiently~\cite{Luo2006NJP8.213}. Conversely, inelastic scatterings, which lead to heating and atomic loss~\cite{Peng2024Commun.Phys.7.101}, should be minimized to maintain the stability of the system~\cite{Rem2013PRL110.163202, Fletcher2013PRL111.125303, Waseem2017PRA96.062704, Kurlov2017PRA95.032710}.  To date, experiments involving the $p$-wave Feshbach resonance in cold Fermi gases have observed inelastic collisions mediated by magnetic dipole-dipole interactions, which typically exhibit a pronounced and narrow atomic loss peak due to energy release~\cite{Zhang2004PRA70.030702R,Schunck2005PRA71.045601,Braun2021PRR3.033269,Ticknor2004PRA69.042712,Nakasuji2013PRA88.012710,Fuchs2008PRA77.053616,Gerken2019RPA100.050701R}. Such inelastic collisions pose a challenge in achieving longer lifetimes and require meticulous attention.

On the other hand, spin-exchange (SE) collisions refer to a category of two-body inelastic interactions mediated by electrostatic interactions~\cite{Ismail2019PHDthesis,Stoof1988PRB38.4688}. These collisions occur when an entrance channel transitions into one or more accessible channels while preserving total angular momentum and orbital angular momentum~\cite{Guo2022PRA105.023313}. SE collisions are significant because electrostatic interactions are often considerably stronger than magnetic dipole interactions, which raises questions about their impact on the system's stability and potential for efficient evaporative cooling to the ultracold regime. 

The strong electrostatic interactions in SE collisions, when combined with the effects of a Feshbach resonance, is particularly interesting for understanding atomic collisions more generally. Particularly, exploring SE collisions near a $p$-wave Feshbach resonance can lead to broader collision widths and potentially weaker peak strengths near the resonance. This interplay between Feshbach resonances and SE collisions is pivotal in determining the stability of ultracold atomic systems and their potential for efficient cooling to the ultracold regime. Exploring these interactions provides deeper insights into the fundamental behaviors and properties of atomic collisions.

In this letter, we report  the observation of a broad state-to-state spin-exchange (SE) collision near a $p$-wave Feshbach resonance at 225 G within a two-component ultracold $^6$Li Fermi gas. 
During this process, two $^6$Li atoms,  initially in the states $|1\rangle\equiv|f=1/2, m_f=1/2\rangle$ and $|3\rangle\equiv|f=3/2, m_{f}=-3/2\rangle$, interact to produce a pair of atoms both in the state $|2\rangle\equiv|f=1/2, m_f=-1/2\rangle$, following the scheme $|1\rangle+|3\rangle\rightarrow|2\rangle+|2\rangle$.
By employing a robust optical dipole trap (ODT), we have successfully isolated the atoms in the $|2\rangle$ state and characterized the time evolution of the collisional process. 
Our findings substantiate the notion that a $p$-wave Feshbach resonance can be significantly altered by an accompanying SE collision, which can modulate the resonance characteristics.
To obtain a more detailed understanding of the $p$-wave resonance identified in this study, we conducted a numerical coupled-channel (cc) analysis specifically designed for the investigation of $p$-wave Feshbach resonances in $^6$Li. The results of this analysis are in good agreement with the experimental data, confirming the theoretical models. Remarkably, the SE collision-influenced $p$-wave Feshbach resonance demonstrated here is characterized by a significantly lower inelastic collision rate and a broader resonance profile compared to those previously reported for $p$-wave resonances in $^6$Li~\cite{Zhang2004PRA70.030702R,Schunck2005PRA71.045601,Fuchs2008PRA77.053616,Gerken2019RPA100.050701R}.
This finding holds considerable potential for advancing the study of SE interactions and opens up novel avenues for the creation of a strongly interacting ultracold $p$-wave Fermi gas.

\begin{figure*}[htbp]
	\begin{center}
		\includegraphics[width=2\columnwidth, angle=0]{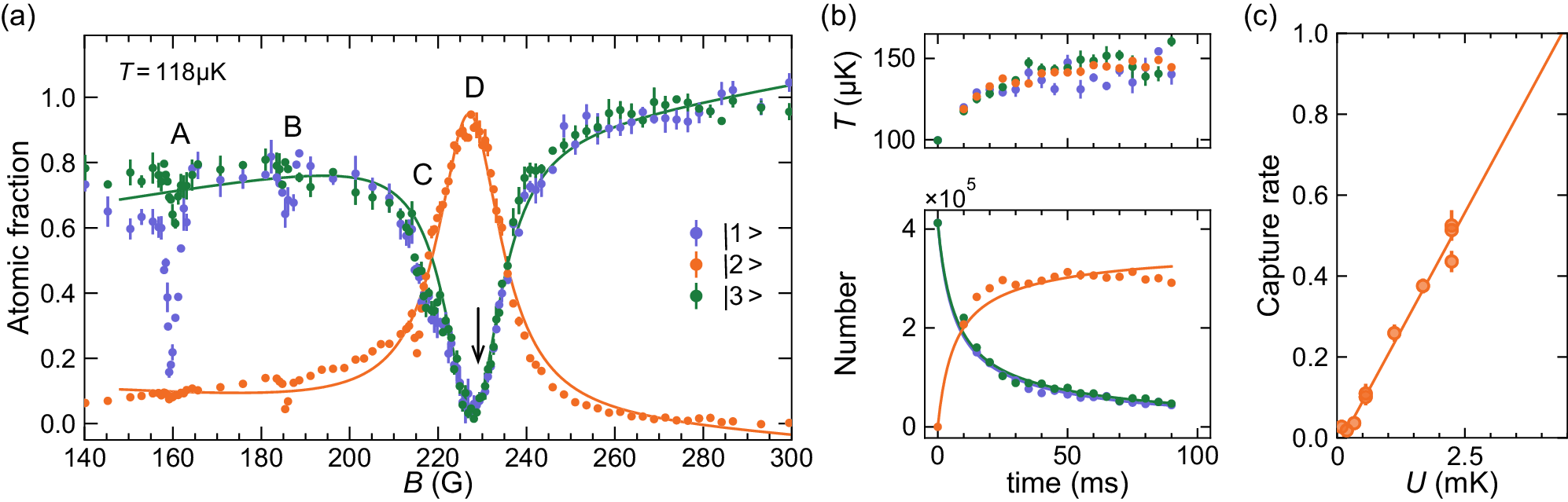}
		\caption{(a) spectrum of residual or gained atomic fraction after a collision time duration of 300 ms. The mixed $|1\rangle$ and $|3\rangle$ gas is prepared in a deep ODT of $U=2.24\ \mathrm{mK}$. The four peaks are labeled from left to right as A to D. The $|2\rangle$ atoms are trapped near resonance D. The solid lines are the best fits using a Lorentzian line-shape combined with a linear slop background. We obtain a full width at half maximum of 15.2(5) G for the $|1\rangle$ and $|3\rangle$ atoms and of 17.5(5) G for the $|2\rangle$ atoms.
		Errorbars are the standard derivation of 2-3 measurements. (b) time evolution of the gas temperature (top) and the atom number (bottom) at the resonance center of D (marked as an arrow in (a)). $T$ increases by 45 $\mu$K for all three states. The solid lines show the fitting results, with $L_2=1.68(3) \times 10^{-16}\ \mathrm{m^3/s}$ for $|1\rangle$ and $|3\rangle$ and  $L^{'}_2=-2.87(5)\times 10^{-16}\ \mathrm{m^3/s}$ for $|2\rangle$.
		(c) capture rate of the $|2\rangle$ atom versus the trap depth. The solid line is a linear fit with a slope of 0.23 mK$^{-1}$. }\label{Fig:MagSweep_AtomLoss}
	\end{center}
\end{figure*}

To achieve a mixture of atoms in the states  $|1\rangle$ and $|3\rangle$, we initially prepared a gas in states $|1\rangle$ and $|2\rangle$ within a cross-beam ODT, following the same procedure described in\cite{Peng2024Commun.Phys.7.101, Li2023arXiv2212.08257}.
Subsequently, we swept the magnetic field from 320 G to 563 G, where the $s$-wave scattering lengths between the states approach zero. This condition facilitated the transition of internal states from $|2\rangle$ to $|3\rangle$ via radio frequency (rf) pulses employing a Landau-Zener passage (LZP). The rf frequency was initiated at 84.052 MHz and linearly ramped to 84.077 MHz over 15 ms, since the Zeeman energy difference between $|2\rangle$ and $|3\rangle$ is 84.066 MHz.
Following the LZP, we selectively removed the remaining atoms in state $|2\rangle$  using a resonant laser pulse, thus obtaining a balanced mixture of atoms in states  $|1\rangle$ and $|3\rangle$, with an  approximate atom number of  3$\times 10^5$ per spin state. To facilitate the trapping of atom in state $|2\rangle$, we incrementally increased the trap depth $U$ from 0.56 mK to 2.24 mK, yielding  a thermal gas with a temperature $T$ of 118 $\mu$K. The resulting trap parameter $\eta=U/k_B T$ was approximately 19~\cite{Li2016PRA93.041401R}, a suitable condition for the efficient confinement of collision products. Here, $k_B$ is the Boltzmann constant.

We conduct a search for the enhanced atom loss by sweeping the magnetic field over a wide range from 140 G to 300 G. The magnetic field is first lowered to 320 G to ensure gas thermalization, and then precisely ramped down to the target field in 20 ms and held for a duration of 300 ms. It is then swept back to 563 G for atom detecting. Using time-of-flight absorption imaging, we record the numbers of remaining atoms $N_{1, 2, 3}$ and the Gaussian widths $\sigma_{x, y, z}$ of the gases in each state. From these $\sigma$ values, we calculate the temperatures.

The atom losses are normalized to the initial atom number of $|3\rangle$ and shown in Fig.\ref{Fig:MagSweep_AtomLoss}(a).
We identify four peaks from the atom losses. 
These peaks, labeled as A, B, and C, occur at magnetic field of 159 G, 185 G, and 215 G, respectively. These correspond to the $p$-wave Feshbach resonances between $|1\rangle$ and $|2\rangle$ in $^6$Li, as inferred from the consistent of the magnetic field regions~\cite{Zhang2004PRA70.030702R, Schunck2005PRA71.045601}. 
In addition, we have detected a broad resonance, labeled D, spanning approximately 20 G around 225 G. 
Notably, an atom increase in the $|2\rangle$ accompanied by the loss of the $|1\rangle$ and $|3\rangle$ atoms near this resonance is also observed. 
Besides its broad width, resonance D displays a symmetric profile, setting it apart from the narrower $p$-wave resonances (A, B, C) at lower magnetic fields, which exhibit significant temperature-dependent broadening at the higher gas temperatures~\cite{Li2018PRL120.193402}.
We fit the data near resonance D using a Lorentzian line shape combined with a linear slop background $\textendash$ caused by the slow magnetic sweep~\cite{Chen2021PRA103.063311} $\textendash$ and determine the resonance center to be at 227.8(1) G for the $|1\rangle$ and $|3\rangle$ atoms and  at 227.1(1) G for the $|2\rangle$ atoms.

We explore the dynamics of the collisions near the resonance D by measuring the temporal evolution of the states, as shown in Fig.\ref{Fig:MagSweep_AtomLoss}(b). We find that the populations of atoms in $|1\rangle$ and $|3\rangle$  are subject to a two-body decay of $d{N_{1, 3}}/dt=-L_2N_{1}N_{3}/V_{eff}$ with an average loss rate of $L_2=1.68(3) \times 10^{-16}\ \mathrm{m^3/s}$.  Here, $V_{eff}=(2\pi)^{3/2}\sigma_x\sigma_y\sigma_z$  represents the average effective volume of the atom clouds in $|1\rangle$ and $|3\rangle$, with their intrap $\sigma_{x, y, z}$ values~\cite{Peng2024Commun.Phys.7.101}. 
At the same time, we detect an increase in the population of atoms in $|2\rangle$, which is resulted from the two-body collision of $|1\rangle$ and $|3\rangle$. This collision is described by a rate equation of $d{N_{2}}/dt=-L^{'}_2N_{1}N_{3}/V_{eff}$ with a rate of $L^{'}_2=-2.87(5) \times 10^{-16}\  \mathrm{m^3/s}$. 
This rate almost aligns with the prediction of an SE collision, where each collision converts a pair of $|1\rangle$ and $|3\rangle$ atoms into a pair of $|2\rangle$ atoms. 
The absolute value of the $L^{'}_2$ is slightly lower than twice the rates of $|1\rangle$ and $|3\rangle$ atoms because some of the $|2\rangle$ atoms leave the trap due to the finite trap depth.
Figure \ref{Fig:MagSweep_AtomLoss}(c) presents the maximum capture rate of the $|2\rangle$ atom, $\Delta N_{|2\rangle}/(\Delta N_{|1\rangle}+\Delta N_{|3\rangle})$, against the trap depth. The data shows a linearly increasing slope of 0.23 mK$^{-1}$, indicating that the atoms in $|2\rangle$ can be fully trapped if $U$ reaches 4.3 mK. 
This estimated value of 4.3 mK is almost ten times of the released energy of the SE collision (which is 24.3 MHz in  $|1\rangle+|3\rangle\rightarrow|2\rangle+|2\rangle$), about 3$k_B\times$388 $\mu$K (average energy of each trapped atom). The trap parameter $\eta$ is closed to 10, a typical value for evaporative process.
As $|2\rangle$ atoms are trapped in our system, there is an associated increase in gas temperatures of 45 $ \mu $K due to the energy release.
This observed temperature increment is lower than the predicted theoretical value, which can be ascribed to the preferential escape of high-energy atoms from the  $|2\rangle$ state. 
The dynamic is illustrated in the upper panel of Fig.~\ref{Fig:MagSweep_AtomLoss}(b). The underlying mechanisms and the quantitative description of this phenomenon can be modeled using the methods in Ref.~\cite{Nuske2015PRA91.043626, Welz2023PRA107.053310}.

Within the mixed Fermi gas comprising states $|1\rangle$ and $|3\rangle$, three $p$-wave scattering channels are identified: $|\{1,1\}\rangle$, $|\{1,3\}\rangle$, and $|\{3,3\}\rangle$.  For clarity, the scattering channel $|\{\alpha,\beta\}\rangle$ is defined based on the atomic basis $|f_1 m_{f1}, f_2 m_{f2}\rangle$.
These channels are subject to two-body inelastic collisions, which may result in decay into various exit channels. In the regime of low collision energy, the exit channels possess lower energy than the corresponding entrance channel.
Since the $|\{1,1\}\rangle$ channel has the lowest energy, it cannot undergo two-body inelastic collisions, so that only inelastic three-body collisions are possible~\cite{Yoshida2018PRL120.133401}. Such collisions manifest as significant atom loss in the state $|1\rangle$ in proximity to resonance A in Fig.\ref{Fig:MagSweep_AtomLoss}(a).
In contrast, the $|\{3,3\}\rangle$ channel has three potential exit channels: $|\{2,2\}\rangle$, $|\{1,3\}\rangle$ and $|\{2,3\}\rangle$. These channels are candidates for atom loss via magnetic dipolar relaxation, particularly above resonance D. However, as illustrated in Fig.\ref{Fig:MagSweep_AtomLoss}(a), no significant atom loss was detected in this regime.

For the $|\{1,3\}\rangle$ channel, the possible exit channels include 
$|\{1,1\}\rangle$, $|\{1,2\}\rangle$ and $|\{2,2\}\rangle$. Transitions to $|\{1,1\}\rangle$ and $|\{1,2\}\rangle$ can occur through dipolar relaxation, while the transition to $|\{2,2\}\rangle$ may proceed via SE collisions. 
Experimental results from our study suggest that SE collisions are the predominant mechanism for atom loss in the $|\{1,3\}\rangle$ channel.  
This predominance is likely attributed to the relatively minor energy difference between the entrance and exit channels in SE collisions, which is significantly lower by a factor of at least three compared to that in dipolar relaxation collisions, thereby resulting in a more robust coupling strength.

\begin{figure*}[htbp]
	\begin{center}
		\includegraphics[width=2\columnwidth,angle=0]{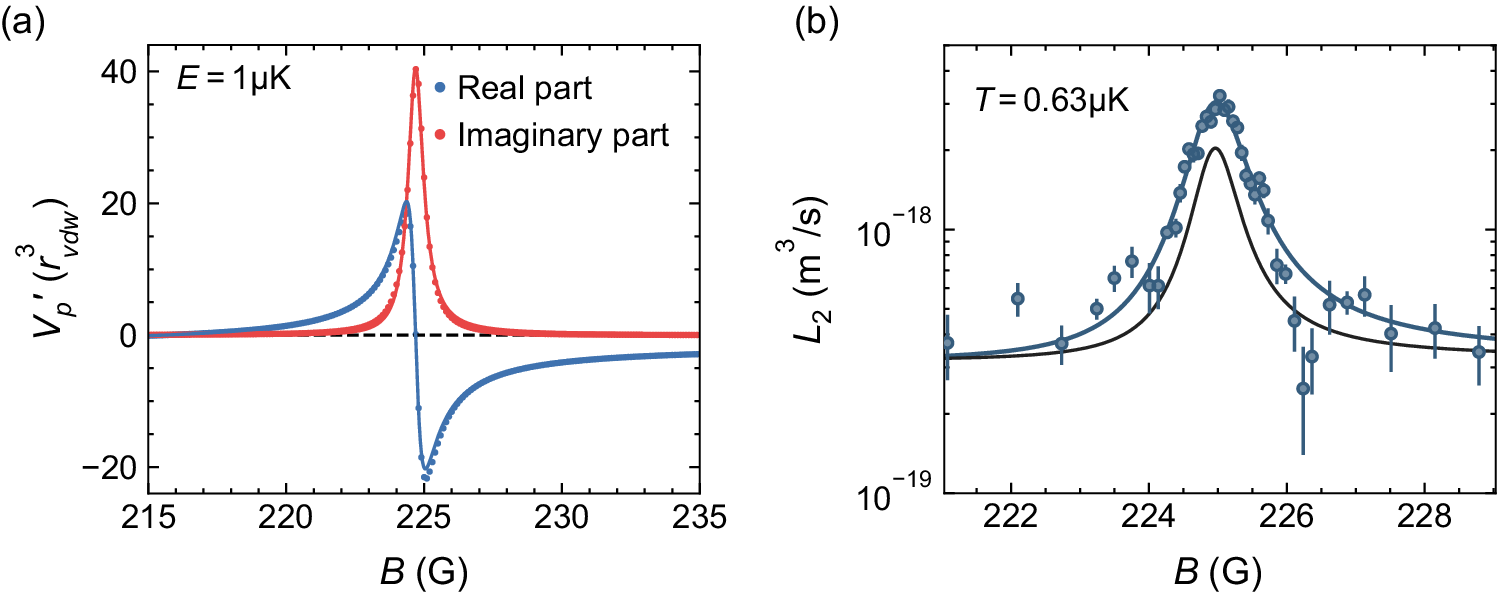}
		\caption{(a) the real (blue dots) and the imaginary (red dots) parts of the scattering volume for the $|1\rangle-|3\rangle$ $p$-wave Feshbach resonance with $E=1\mathrm{\mu K}$. The lines represent their best fit results. (b) the two-body loss rate $L_2$ in terms of magnetic field when $T$= 0.63 $\mathrm{\mu K}$. The blue line represents the fit of Eq.~\ref{eq:L_2} with free parameters $V_1,\ B_0$ and $T$, while the black line shows the calculated result with the theoretical value $V_1$ from the cc calculation. A offset loss rate of $3.2(5)\times 10^{-19}\ \mathrm{m^3/s}$ is added in the fits to account the atom loss from the evaporative cooling.
		}\label{Fig:Resonant_parameters}
	\end{center}
\end{figure*}

To obtain more collision properties of this observed broad resonance, we conduct a numerical calculation of the cc model, specific to $^6$Li $p$-wave resonances. Overall, the cc calculations confirm that the observed resonance D near 225 G is a $p$-wave Feshbach resonance since the experiments agree very well with the theoretical predictions. 
We also refer to this broad resonance as $|1\rangle$- $|3\rangle$ $p$-wave Feshbach resonance, reflecting the involved entrance channel.

A detailed theoretical framework for this cc model, including the effective singlet and triplet interaction potentials, is also presented in our previous work~\cite{Zhang2022CPB31.063402,Li2021JOPB54.115201}. Here, we only provide a brief overview. 
The model begins with a Hamiltonian of two colliding atoms subjected to an external magnetic field $B$ and an interatomic separation $R$, expressed as $	H=H_0+V_Z(R)+V_{hf}(R)+V_d(R)$.
The $H_0$ encompasses the relative motion, the $p$-wave centrifugal barrier, and the interaction potential of the two colliding atoms in the absence of external fields. $V_Z(R)+V_{hf}(R)$  denote the Zeeman interaction and the hyperfine potential. $V_d(R)$ denotes the magnetic dipole-dipole interaction potential.
We obtain the cc equations by substituting $H$ into a time-independent Schr\"{o}dinger equation, and we solve these equations using the log-derivative method to get the scattering matrix $S(k)$~\cite{Manolopoulos1986J.Chem.Phys.85.6425}, where $k$ is the wavevector in the entrance channel.
As the threshold energy of the entrance channel $|\{1,3\}\rangle$ is set to 0, the collision energy $E=\hbar^2 k^2/2\mu$, where $\mu$ is the reduced mass of $^6$Li atom pair, and $\hbar$ is the reduced Plank constant.

In the presence of inelastic collisions, the diagonal $S$-matrix element of the entrance channel has magnitude $|S_{\{|\{1,3\}\rangle,|\{1,3\}\rangle\}}(k)|\leq 1$~\cite{Hutson2007NJP9.152}, and the scattering volume become complex with a equation of  
\begin{equation}
	V^{'}_p(k)=\frac{1}{ik^3}\frac{1-S_{\{|\{1,3\}\rangle,|\{1,3\}\rangle\}}(k)}{1+S_{\{|\{1,3\}\rangle,|\{1,3\}\rangle\}}(k)}.
\end{equation}
In Fig.~\ref{Fig:Resonant_parameters}(a), we plot the calculated real part, $\Re[V^{'}_p]$, and the imaginary part, $\Im[V^{'}_p]$, of the $|1\rangle$- $|3\rangle$ resonance with $E=1\ \mathrm{\mu K}$.
It can be seen that inelastic collision smooths the poles of elastic scattering.

In the current cc model, the magnetic dipole-dipole interaction has been omitted. This is because the energy splitting between the different $m_l$ components due to the dipole-dipole interaction can be independently estimated, and it is found to be approximately 10 mG for the $p$-wave resonances in $^6$Li~\cite{Zhang2022CPB31.063402,Chevy2005PRA71.062710}. When comparing the ten Gaussian widths of the $V^{'}_p$, such negligible splitting can therefore be ignored.

To analyze the effects of inelastic collisions within our model, we introduce an imaginary component of $1/V^{'}_p(k)\rightarrow1/V_p(k)+i/V_1$~\cite{Kurlov2017PRA95.032710,Naik2011EPJD65.55}, where $1/V_p(k)=1/V_{bg}[1-\Delta B/(B-B_0)]+k_e k^2$ is the inverse of the energy-dependent scattering volume $V_p(k)$ under an effective range expansion, and the energy-independent parameter $V_1$ denotes the width and strength of the inelastic collisions.  $V_p(k)$ also denotes the scattering volume of the system in absence of inelastic scattering.
By fitting $V^{'}_p$, we have determined the resonance center $B_0=224.693$ G, the resonance width $\Delta B=-9.22$ G, the background scattering volume $V_{bg}=-4.58\times 10^4\ a^3_0$, and $V_1=1.84\times 10^{-25}\ m^3$. 
These values yield a $p$-wave resonance strength parameter $s_{res}$~\cite{Dong2016PRA94.062702} of $ 0.05$, which is comparable to the values observed for other $p$-wave resonances in $^6$Li, and the value of $B_0$ agrees with the observed value in Fig.\ref{Fig:MagSweep_AtomLoss}(a).
In this fit, we define the effective range term $k_e$ as $\hbar^2/(mV_{bg}\delta \mu\Delta B)$, where $\delta \mu=120\ k_B\times \mathrm{\mu K/G}$ represents the magnetic moment difference between the molecular channel and entrance channel as calculated by cc method, which accounts for the energy shift of the peak value in $\Im[V^{'}_p]$.
It is worth to note that the $V_1$ for the $|1\rangle$- $|3\rangle$ resonance is about 4 orders of magnitude smaller than the value found in $|1\rangle$-$|2\rangle$ $p$-wave resonances, which was measured to be 4.85 $\times 10^{-21}$m$^{3}$ in Ref.~\cite{Waseem2017PRA96.062704}.

Incorporating the effects of inelastic scattering, we also modify the scattering amplitude $f_p(k)$ and the inelastic collision rate $\beta_{in}(k)$, which are
\begin{equation}\label{eq:Scattering_amplitude}
	f_p(k)=\frac{-k^2}{1/V_p(k)+ik^3+i/V_1},
\end{equation}
and 
\begin{equation}\label{eq:inelastic_loss_rate}
	\beta_{in}(k)=\frac{24\pi\hbar}{mV_1}\frac{k^2}{[1/V_p(k)]^2+[1/V_1+k^3]^2},
\end{equation}
respectively.
For a thermal gas, the two-body atom loss rate $L_2$ is then obtained by performing an average of $\beta_{in}(k)$ over the Boltzmann distribution as follows
\begin{equation}\label{eq:L_2}
	L_2=\frac{2}{\sqrt{\pi}(k_BT)^{3/2}}\int_{0}^{\infty}dE\sqrt{E}\text{Exp}\left(-\frac{E}{k_BT}\right)\beta_{in}(E)
\end{equation}
In this experiment, we have measured $L_2(B)$ of the $|3\rangle$ atoms near the $|1\rangle$- $|3\rangle$ resonance with a gas temperature of 0.63 $\mathrm{\mu K}$, as shown in Fig.~\ref{Fig:Resonant_parameters}(b). 
We fit the data with Eq.\ref{eq:L_2} and get values of $V_1=1.4(1)\times 10^{-25}$ m$^{3}$, $T=1.3(1)\ \mu$K, and $B_0=224.951(15)$ G. Thus, the values of $V_1$ are consistent both theoretically and experimentally. 
Moreover, due to the much smaller energy shift than that observed in Fig.\ref{Fig:MagSweep_AtomLoss}(a), the obtained resonance center in this measurement is in a very good agreement with the theoretically predicted value of the cc model, which confirms the reliability of our calculations.

\begin{figure}[htbp]
	\begin{center}
		\includegraphics[width=\columnwidth, angle=0]{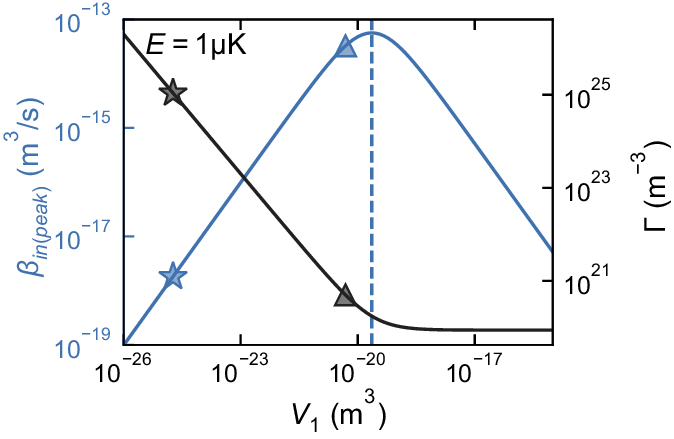}
		\caption{The peak(blue line) and width(black line) of $\beta_{in}$ at various $V_1$. The vertical dashed line indicates the turning point of $V_1\equiv 1/k^3$. The values of $|1\rangle$-$|3\rangle$ resonance are marked as the stars, and those of $|1\rangle$-$|2\rangle$ $p$-wave resonance are marked as the triangles.
		}\label{Fig:betain_peak_and_width}
	\end{center}
\end{figure}

Equation~\ref{eq:inelastic_loss_rate} illustrates that $V_1$ plays an important role in determining both the width $\Gamma$, which is $\propto 2(1/{V_1}+k^3)$, and the peak inelastic collision rate $\beta_{in(peak)}(k)$, which is $\propto k^2/(1/V_1+V_1 k^6+2k^3)$, of the inelastic collisions. 
We plot $\Gamma$ and $\beta_{in(peak)}(k)$ at various $V_1$ in Fig.~\ref{Fig:betain_peak_and_width} when $E=1\mathrm{\mu K}$. It shows that $\beta_{in(peak)}(k)$ increases with $V_1$ and starts to decrease upon reaching the turning point $V_1\equiv 1/k^3$. Simultaneously, $\Gamma$ decreases with $V_1$ increases and saturates at $2k^3$.
When the gas temperature reaches a low value, the much smaller value $V_1$ of the $|1\rangle$-$|3\rangle$ $p$-wave Feshbach resonance (marked as stars) induces a much weaker and broad inelastic loss  compared to the other $p$-wave Feshbach resonances in $^6$Li (marked as triangle), which has much higher value of $V_1$.

In conclusion, we have studied the properties of two-body state-to-state spin-exchange collision near the $|1\rangle$-$|3\rangle$ $p$-wave Feshbach resonance of $^6$Li both experimentally and theoretically.
The experimental outcomes underscore the pivotal role of SE collisions in shaping the decay processes, as evidenced by our successful capture and detection of $|2\rangle$ atoms within a deep optical dipole trap.
The congruence between the experimental data and the predictions of coupled-channel models for the Feshbach resonance is important, particularly given the broad resonance range and the subdued rate of inelastic collisions observed. 
These insights have the potential to significantly advance evaporative cooling techniques and foster the creation of a durable $p$-wave unitary Fermi gas~\cite{Top2021PRA104.043311}.

Notably, the signal of atom loss near 225 G was also mentioned in the studies of $^6$Li three-component gases~\cite{Lompe2011PHDthesis,HuckansPRL102.165302}. However, this work represents the first comprehensive experimental and theoretical investigation of the $|1\rangle$-$|3\rangle$ $p$-wave Feshbach resonance.

Moreover, the spectral analysis presented in Fig.\ref{Fig:MagSweep_AtomLoss}(a) discloses a more intricate resonance structure than initially anticipated. Deviations from predicted loss patterns were observed; notably, a minor loss of atoms in state $|3\rangle$ occurred at resonance A, and a comparable loss of atoms in states  $|1\rangle$ and $|3\rangle$ at resonance C.
These unexpected losses may be indicative of three-body collisions involving atoms in all three states~\cite{Schumacher2023arXiv2301.02237}. Future investigations are warranted to unravel the complexities of these phenomena.

\textit{Acknowledgements}
 This work receives support from the National Natural Science Foundation of China under Grant No.12174458 and 11804406 and from Fundamental Research Funds for the Central Universities, Sun Yat-sen University under Grant No. 2021qntd28. 
%
	

\end{document}